# SEARCH FOR A SIGNAL ON PHASE TRANSITIONS OF STRONGLY INTERACTING MATTER USING THE NUCLEAR TRANSPARENCY EFFECT

## M. Ajaz[1], M.K.Suleymanov[1,2], Ali Zaman[1], K.H. Khan[1], Z.Wazir[1]

[1] *Department of Physics, COMSATS Institute of Information Technology, Islamabad, Pakistan*

[2] *Veksler and Baldin Laboratory of High Energy Physics, JINR, Russia*

## Abstract

We discuss that the results coming from the central experiments confirm the results which had been obtained for the behavior of the $K^+$-meson's temperature behavior as a function of the energy in SPS energy range. To see the "horn" for the behavior of the ratio for average values of $K^+$- to $\pi^+$- mesons as a function of centrality the new more rich experimental data are required. The data can be obtained with NICA/MPD setup. The existing of the QCD critical point could be identified by using the nuclear transparency effect as a function of the centrality.

## 1. Introduction.

Ultrarelativistic heavy ion collisions provide a unique opportunity to create and study the nuclear matter at high densities and temperatures. The produced state will pass different phases of the strongly interacting matter (SIM). There are some effects which are considered as possible signatures on strongly interacting matter as well as the Quark Gluon Plasma formation.

## 2. Ga'zdzicki's key results on new phases of strongly interacting matter.

Among all other results on ultrarelativistic heavy ion collisions concerning the states of the SIM, the results which were obtained by M. Ga'zdzicki [1]–were more attractive. The paper [1] discussed that the results [2-3] on the energy dependence of hadron production in central *Pb+Pb* collisions at 20, 30, 40, 80 and 158 A GeV coming from the energy scan program at the CERN SPS serve as evidence for the existence of a transition to a new form of strongly interacting matter, the Quark Gluon Plasma (QGP) [4]. Thus they are in agreement with the conjectures that at the top SPS and RHIC energies the matter created at the early stage of central *Pb+Pb* and *Au+Au* collisions is in the state of QGP [5-6]. The key results are summarized in Fig. 1. The most interesting effect can be seen in the energy dependence of the ratio $<K^+>/<\pi^+>$ of the mean multiplicities of $K^+$ and $\pi^+$ produced per event in the central *Pb+Pb* collisions, which is plotted in the top panel of the figure. Following a fast threshold rise, the ratio passes through a sharp maximum in the SPS range and then seems to settle to a lower plateau value at higher energies. Kaons are the lightest strange hadrons and $<K^+>$ is equal to about half of the number of all anti-strange quarks produced in the collisions. Thus, the relative strangeness content of the produced matter passes through a sharp maximum at the SPS in nucleus-nucleus collisions. This feature is not observed for proton-proton reactions.

A second important result is the constant value of the apparent temperature of $K^+$ mesons in central *Pb+Pb* collisions at low SPS energies as shown in the bottom panel of the figure. The plateau at the SPS energies is preceded by a steep rise of the apparent temperature at the AGS and followed by a further increase indicated by the RHIC data.

Very different behaviour is measured in proton-proton interactions. Presently, the sharp maximum and the following plateau in the energy dependence of the <$K^+$>/<$\pi+$> ratio has only been reproduced by the statistical model of the early stage [4] in which a first order phase transition is assumed. In this model the maximum reflects the decrease in the number ratio of strange to non-strange degrees of freedom and changes in their masses when deconfinement sets in. Moreover, the observed steepening of the increase in pion production is consistent with the expected excitation of the quark and gluon degrees of freedom. Finally, in the picture of the expanding fireball, the apparent temperature is related to the thermal motion of the particles and their collective expansion velocity. Collective expansion effects are expected to be important only in heavy ion collisions as they result from the pressure generated in the dense interacting matter. The stationary value of the apparent temperature of $K^+$ mesons may thus indicate an approximate constancy of the early stage temperature and pressure in the SPS energy range due to the coexistence of hadronic and deconfined phases, as in the case of the first order phase transition [7-8].

Thus, the anomalies in the energy dependence of hadron production in central *Pb+Pb* collisions at the low SPS energies serve as evidence for the onset of deconfinement and the existence of QGP in nature. They are consistent with the hypotheses that the observed transition is of the first order. The anomalies are not observed in p+p interactions and they are not reproduced within hadronic models [9].

These results and their interpretation raise questions which can be answered only by new measurements.

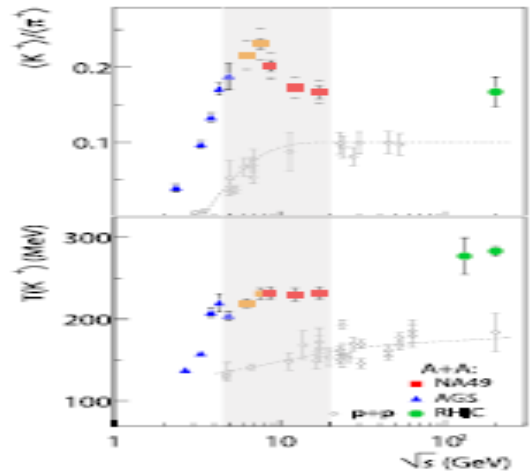

**Fig 1.** Collision energy dependence of the $K^+$ to $\pi^+$ ratio and the inverse slope parameter of the transverse mass spectra measured in central Pb+Pb and Au+Au collisions (solid symbols) compared to results from p+p reactions (open dots). The changes in the SPS energy range (solid squares) suggest the onset of the deconfinement phase transition. The energy region covered by the future measurements at the CERN SPS is indicated by the gray band.

## 3. Central experiment's results

What about the results coming from the central experiments?  Could these experiment indicate the effects same with Ga'zdzicki's key results? Let us now consider some results from the central heavy ion collisions.

During the last several years some results of the central experiments (see for example [10]) are discussed. These results demonstrate the point of regime change and saturation on the behavior of some characteristics of the events as a function of the centrality. One can believe that such phenomena connected with fundamental properties of the strongly interacting mater and could reflect the changes of its states (phases).

In Ref. [11] the variations of average transverse mass of identified hadrons with charge multiplicity have been studied for AGS, SPS and RHIC energies (Fig.2). A plateau was observed in the average transverse mass for multiplicities corresponding to SPS energies. It was claimed that it can be attributed to the formation of a coexistence phase of quark gluon plasma and hadrons. So one can say that the central experiments confirms existing the plateau for the behaviours of *K*-mesons' temperature as a function of  collisions centrality at the SPS energies - the second key result of Ga'zdidcki [1].

Emission of $\pi^{\pm}$, $K^{\pm}$, $\varphi$, and $\Lambda$ was measured in near-central *C+C* and *Si+Si* collisions at 158 *A*GeV beam energy[12]. Together with earlier data for *p+p*, *S+S,* and *Pb+Pb*, the system-size dependence of relative strangeness production in nucleus-nucleus collisions are shown in Fig.3. Its fast rise and the saturation observed at about 60 participating nucleons can be understood as the onset of the formation of coherent systems of increasing size. So we could see that the results coming from the central experiments confirmed a fast threshold rise at AGS energy range.

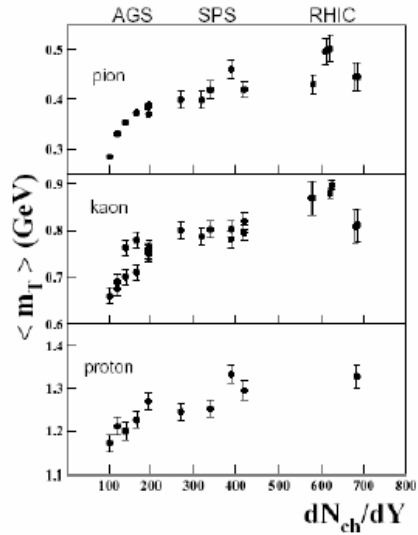

**Fig. 2.** Variation of $<m_T>$ with produced charged particles per unit rapidity at mid rapidity for central collisions corresponding to different √s spanning from AGS to RHIC.

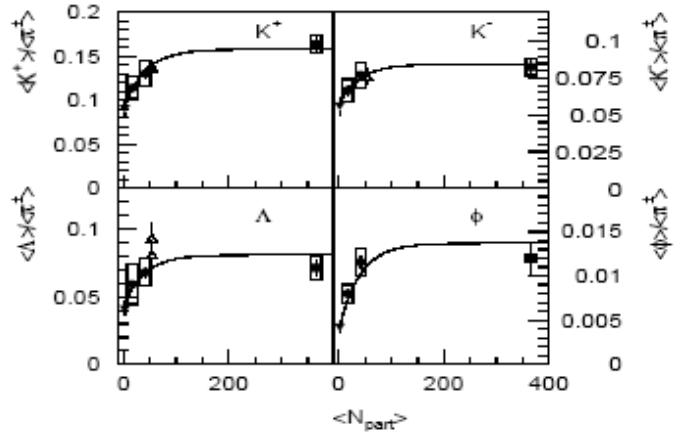

**Fig. 3** The experimental ratios of the average values of multiplicity of $K^+-$, $K^--$, $\varphi -$ mesons and $\Lambda$-hyperons to the average values of multiplicity of $\pi^{\pm}$- mesons as a function of centrality

But these results could not indicate any sharp maximum in the SPS range.
So we could say that:
- The centrality experiments confirms the existing the plateau for the behaviours of $K$-mesons temperature as a function of collisions centrality at the SPS energies - the second key result of Ga'zdzidcki [1] ;
- These experiments could indicate the increasing of the ratio $<K^+>/<\pi^+>$ at AGS energies - the first key result of Ga'zdzidcki, but could not show the sharp maximum in the SPS range.

We think that the last result could be connected with poorness of the experimental data for the ratio $<K^+>/<\pi^+>$ around of the values $N_{part} \sim 60$. This area could be investigated intensity by NICA [13] and CBM [14] experiments.

## 4. Nuclear Transparency Effect.

New phases of strongly interacting matter (information on the QCD critical point) could be identified using the nuclear transparency effect - the behavior of $R=a_{AA}/a_{NN}$ function at different energies as a function of the centrality; because the transparency capability of different states of nuclear matter must be different.
We believe that the study of the behavior of $R$ at different energies as a function of the centrality could give the information on the onset state of the deconfinement as well as on QCD critical point. Using data coming from codes and experimental data on the behavior of $R$ it is possible to get information on the appearance of the anomalous nuclear transparency.
Nuclear transparency is one of the effects of nuclear-nuclear collision from which one may get the information about the structure, states, properties and phases of the nuclear matter. The stated effect is a promising observable to map the transition between the different states/phases of the nuclear medium to the propagation of hadrons. Transparency depends upon different factors of the collisions.
Now one need to address the important question that how to fix the centrality and how one can study the centrality dependency of the $R$ in the experiment?

In papers [15] the nuclear transparency was used for the analysis of the data coming from the hadron- nuclear experiments. For this aim the authors used the inner cone term which was defined from the *NN* interaction (see Fig. 4). The average values of multiplicity of fast particles – *s*-particles ($n_s$) were defined in the cone with half angle $\theta$. Then the average values of *s*-particles with emitted angle less then $\theta$ were defined in nucleon-nuclear interactions (*NA*) and studied as a function of the number of the heavy particles - *h*-particles ($N_F$) emitted in the interactions.

In the Ref. [16], the effect of "transparency" of nuclear matter in interactions between $\pi^-$ - mesons and carbon nuclei was investigated at $P_{\pi^-}=40$ *GeV/c*. The following are their findings: for all chosen values of the limiting emission angle $\theta_n$ ($2.5^0$, $8^0$, and $10^0$) the average multiplicity of the $\pi^{\pm}$- mesons of the inner cone does not depend on the number of emitted protons ($N_p$), and for $\theta_n =2.5^0$ and $8^0$ it coincides with the results for the $\pi^-p$-interactions; the fact that the average multiplicity of the $\pi^+$- mesons of the inner cone is independent of the number of protons ($N_p$) emitted from the carbon nucleus does not mean total "transparency" of the nucleus to these particles, since their average energy decreases with increasing ($N_p$) (see Fig.5).

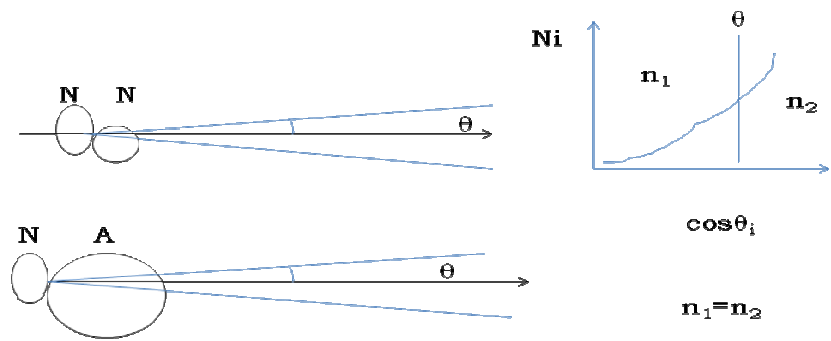

**Fig.4**

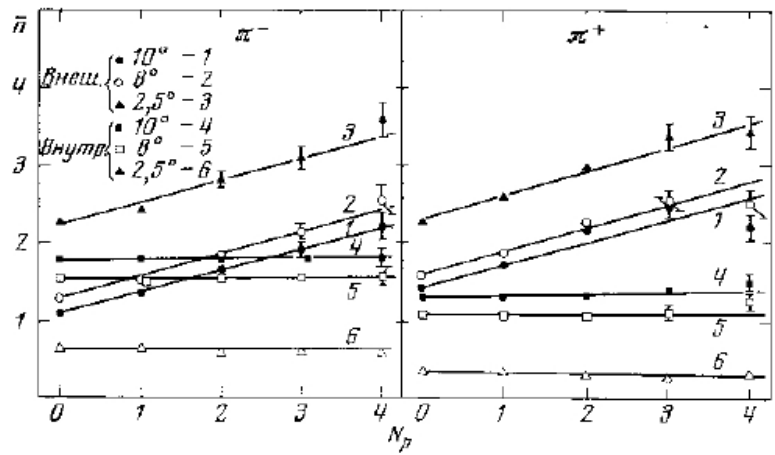

**Fig.5** The average values of multiplicity for $\pi^-$ (left panel) and $\pi^+$ (right panel) – mesons as a function of a number of identified protons in $\pi^-{}^{12}C$-reactions (lines were drift by hand)

## Conclusion

The results coming from the central experiments confirm the existing of the saturation for the behaviours of *K*-mesons' temperature as a function of collisions centrality at the SPS energies.

These experiments could indicate the increasing of the ratio $<K^+>/<\pi^+>$ at AGS energies, but could not show the sharp maximum in the SPS range. The result could be connected with insufficiency of the experimental points in the region of sharp maximum ($N_{part}$~60).

NICA and FAIR experiments could get the necessary data to cover the region $N_{part}$~60.

Study of the nuclear transparency effect as a function of the centrality could give an important information of the phases of strongly interacting matter.

We offer to use the inner cone definition to study the nuclear transparency effect as a function of the centrality.